\documentclass[12pt]{JHEP3}
\usepackage{amsmath, amssymb}
\usepackage{epsfig}
\usepackage{soul}

\newcommand{\beq}{\begin{equation}}
\newcommand{\eeq}{\end{equation}}
\newcommand{\baq}{\begin{eqnarray}}
\newcommand{\eaq}{\end{eqnarray}}
\newcommand{\mc}[1]{\mathcal{#1}}

\def\N{{\bar{N}}}

\def\l{\lambda}

\def\bpo{|\bar{\phi}_0|}
\def\vp{|\varphi|}

\def\bvpo{|\bar{\varphi}_0|}

\def\ä{\"{a}}

\def\p{|\phi|}
\def\po{|\phi_0|}

\def\E{Eq.~}

\title{K\"{a}hler potentials for the MSSM inflation and the spectral index}

\author{ Sami Nurmi~\footnote{E-mail: sami.nurmi@helsinki.fi}~
\\
Department of Physical Sciences, P.O. Box 64, FIN-00014 University
of Helsinki, Finland,\\
Helsinki Institute of Physics, P.O. Box 64, FIN-00014 University of
Helsinki, Finland}

\abstract{

Recently it has been argued that some of the fine-tuning problems of
the MSSM inflation associated with the existence of a saddle point
along a flat direction may be solved naturally in a class of
supergravity models. Here we extend the analysis and show that the
constraints on the K\"{a}hler potentials in these models are
considerably relaxed when the location of the saddle point is
treated as a free variable. We also examine the effect of
supergravity corrections on inflationary predictions and find that
they can slightly alter the value of the spectral index. As an
example, for flat direction field values $\bvpo=1\times10^{-4}M_P$
we find $n\sim0.92~...~0.94$ while the prediction of the MSSM
inflation without any corrections is $n\sim0.92$.

}

\preprint{HIP-2007-54/TH}


\maketitle

\begin{document}


\section{Introduction}

Recently it has been argued that inflation can be realized already
within the Minimally Supersymmetric Standard Model (MSSM)
\cite{AEGM,AEGJM}. In this case the inflaton field is a particular
gauge invariant combination of squarks and sleptons corresponding to
a flat direction\footnote{For a discussion of the properties of flat
directions see e.g. \cite{DRT,GKM} and for a review of their
cosmological implications, e.g. \cite{KARI-REV}.} of the MSSM. Its
couplings to other MSSM degrees of freedom are thus fully determined
and at least in principle measurable in laboratory experiments such
as LHC or a future Linear Collider. This is in sharp contrast with
the conventional models where the inflaton field is usually taken to
be some ad hoc gauge singlet.

As discussed in \cite{AEGM}, the phenomenologically acceptable
candidates for the inflaton field are the dimension six flat
directions {\bf udd} and {\bf LLe}. The potential along these flat
directions can be written to leading order order as
\beq
\label{leading_potential}
  V_0=\frac{1}{2}m^2\p^2-\frac{A\l}{6}\p^6+{\l}^2\p^{10}\ ,
\eeq
where $\p$ is the absolute value of the field parameterizing the
flat direction. Here and elsewhere in the text we use units where
$M_P\equiv1$. The parameters $m$ and $A$ are supersymmetry (SUSY)
breaking terms depending on the underlying supergravity (SUGRA)
model and $\lambda$ is an effective coupling constant associated to
the non-renormalizable operator lifting the flat direction.

For generic values of the SUSY breaking parameters, the potential
Eq.~(\ref{leading_potential}) does not give rise to inflation.
However, if one imposes the condition
\beq
\label{saddlepoint}
  A^2=40m^2 \ ,
\eeq
the potential has a saddle point at
\beq
\label{po}
  \po=\Big(\frac{m}{\sqrt{10}\l}\Big)^{1/4}\ll1\ .
\eeq
Close to the saddle point the potential becomes flat enough to
support inflation and can be expanded as
\beq
\label{Vexp}
  V_0\approx\frac{4}{15}m^2\po^2+\frac{16}{3}\frac{m^2}{\po}(\p-\po)^3\
  .
\eeq
If the initial conditions are such that $\phi\simeq \phi_0$, there
follows a period of slow roll inflation with a very low scale
$H_{\rm inf}\sim 1-10$~GeV producing primordial perturbations at the
observed level and with the spectral index $n\simeq 0.92$
\cite{AEGM}.

The success of the MSSM inflation obviously relies on the existence
of the saddle point. Due to the exceptionally low inflationary
scale, the potential needs to be extremely flat to produce large
enough primordial perturbations. Consequently, the saddle point
condition \E(\ref{saddlepoint}) must be satisfied with an accuracy
of about $10^{-18}$ \cite{AEGJM}. However, as proposed in \cite{EMN}
this apparent fine-tuning problem can be solved
naturally\footnote{It should be kept in mind though that in the MSSM
inflation \cite{AEGM}, the flat direction is the only dynamical
degree of freedom during inflation and the moduli fields of the
underlying supergravity model are thus implicitly assumed to be
stabilized before the beginning of inflation. This represents a
non-trivial constraint in any realistic supergravity model and might
also be a source of additional fine-tuning, see e.g. \cite{Lalak}.}
in a class of supergravity models where the K\"{a}hler potential is
chosen in such a manner that the saddle point condition
Eq.~(\ref{saddlepoint}) is identically satisfied.  In \cite{EMN} it
was found that this can be achieved with K\"{a}hler potentials that
up to quadratic part in $\p$ have a fairly natural form encountered
in various string theory compactifications but that also require
fixing of some higher order terms. In this work we show that the
constraints on the K\"{a}hler potentials are considerably relaxed if
the location of the saddle point is treated as a free variable. In
particular, we find that in order to identically produce the flat
potential required by the MSSM inflation, the K\"{a}hler potential
needs to be completely fixed only up to quadratic terms in $\p$ and
not to higher orders as in \cite{EMN}. This considerably extends the
class of allowed K\"{a}hler potentials and consequently increases
the possibility to find theoretically motivated models that could
yield the MSSM inflation.

We also discuss the effect of supergravity corrections on
inflationary predictions. Although the corrections are suppressed by
powers of $\p$, they become significant in the vicinity of the
saddle point Eq.~(\ref{saddlepoint}) where the first and second
derivative of the leading order potential
Eq.~(\ref{leading_potential}) vanish. We find that in supergravity
models where the MSSM inflation can be naturally realized, the
relevant supergravity corrections to the inflaton potential will
manifest themselves as additional linear terms in  Eq.~(\ref{Vexp}).
This kind of corrections to the MSSM inflation have been discussed
in \cite{LythDimopoulos, AM} (see also \cite{Allahverdi:2007vy} for
a discussion on dark matter and the MSSM inflation) without any
particular supergravity motivation and it is well known that they
can affect the spectral index and total number of e-foldings. The
difference here is that since the corrections arise from a given
supergravity model they are not arbitrary but can be exactly
calculated.

We find that the supergravity corrections generically tend to
increase the value of the spectral index. For the most typical field
values \cite{AEGM} of the MSSM inflation $\bvpo\sim10^{-4}$, where
$\bvpo$ denotes value of the canonically normalized field at the
saddle point, it is fairly easy to find K\"{a}hler potentials that
bring the spectral index close to the observationally favoured value
$n=0.948\pm0.015$ \cite{data}. For smaller field values,
$\bvpo\lesssim10^{-5}$ the corrections become negligible and one
recovers the result $n\sim0.92$ whereas large field values
$\bvpo\gtrsim10^{-3}$ typically yield too large spectral index.
However it is still possible to choose the K\"{a}hler potential such
that the resulting spectral index is consistent with observations
even with large field values.


\section{The supergravity models}

In \cite{EMN} it was found that the saddle point condition of the
MSSM inflation can be satisfied identically in supergravity models
with F-term supersymmmetry breaking and K\"{a}hler potentials of the
form
\baq
\label{fullkahler}
K&=&\sum_m\beta_m{\rm
ln}(h_m+h_m^{*})+\kappa\prod_m(h_m+h_m^{*})^{\alpha_m}\p^2+\mu\Big(\kappa\prod_m(h_m+h_m^{*})^{\alpha_m}\Big)^2\p^4
+\nonumber\\&&
\nu\Big(\kappa\prod_m(h_m+h_m^{*})^{\alpha_m}\Big)^3\p^6+\mc{O}(\p^8),\
\eaq
where $h_m$ denote hidden sector fields and
$\kappa,\beta_m,\alpha_m,\mu,\nu$ are constants. The superpotential
is taken to be of the form
\beq
\label{w}
W=\hat{W}(h_m)+\frac{\hat{\lambda}_6(h_m)}{6}\phi^6,
\eeq
and the hidden sector dependent parts are treated as constants. The
MSSM inflation is not a generic outcome of all such supergravity
models, though, but one needs to place constraints on the parameters
of the K\"{a}hler potential, see \cite{EMN}. Moreover, the hidden
sector or moduli fields $h_m$ need to be stabilized before the
beginning of inflation by some mechanism not consistently taken into
account here.

It turns out that the fairly strict constraints on the K\"{a}hler
potentials found in \cite{EMN} can be considerably relaxed by
allowing the location of the saddle point $\po$ to slightly vary
from the value given by Eq.~(\ref{saddlepoint}). This is indeed a
natural thing to do since Eq.~(\ref{saddlepoint}) results from the
leading order part of the inflaton potential alone while
supergravity models typically yield higher order corrections as
well. Therefore we write $\po$ as an expansion
\beq
\label{fiinolla}
\po^4=\bar{\po}^4(1+\Delta_1\po+\Delta_2\po+...)\ ,
\eeq
where $\bar{\po}$ denotes the leading order part determined by
\E(\ref{saddlepoint}) and the terms
$\Delta_n\po\sim\mc{O}(\bar{\po}^{2n})$ represent yet unfixed higher
order degrees of freedom corresponding to the higher order terms in
the potential.

Using Eq.~(\ref{fiinolla}) and repeating the analysis of \cite{EMN}
one finds that the flat potential of the MSSM inflation is
identically obtained if the parameters in the K\"{a}hler potential
Eq.~(\ref{fullkahler}) are chosen according to Table $(1)$.
\begin{table}[h!]
\begin{center}
\label{table}
\caption{The constraints on the parameters of the K\"{a}hler
potential Eq.~$(2.1)$ implied by the flatness of the inflaton
potential.}
\vspace{12pt}
\begin{tabular}{|c||c||c||c|}
\hline
$\rule{0pt}{3ex}\beta=\sum\beta_m$ & $\alpha=\sum\alpha_m$ &
$\gamma=\sum\frac{\alpha_m^2}{\beta_m}$ &
$\delta=\sum\frac{\alpha_m^3}{\beta_m^2} $\\
\hline
$\rule{0pt}{3ex}-7$ & $~~0$ & $ \frac{1}{4}-3\mu $ & $ \delta $\\
\hline
$\rule{0pt}{3ex}-7$ & $-\frac{25}{9}$ &
$-\frac{46}{81}-\frac{22}{9}\mu$ & $-\frac{2414}{16767}-\frac{628}{1863}\mu-\frac{2804}{207}\mu^2+\frac{162}{23}\nu$\\
\hline
$\rule{0pt}{3ex}-11$ & $-\frac{1}{9}$ &
$\frac{28}{81}-\frac{26}{9}\mu$ & $\frac{6556}{69255}-\frac{3736}{7695}\mu-\frac{12596}{855}\mu^2+\frac{162}{19}\nu$\\
\hline
$\rule{0pt}{3ex}-11$ & $-4$ & $-\frac{7}{8}-\frac{5}{2}\mu$ & $ -\frac{339}{1600}-\frac{73}{200}\mu-\frac{1371}{100}\mu^2+\frac{36}{5}\nu$\\
\hline
\end{tabular}
\end{center}
\end{table}
The conditions in Table $(1)$ are much less restrictive than those
found in \cite{EMN}.\footnote{Note that in Table $(1)$ we have not
included the case with $\beta=-3$ and $\alpha=-4/9$ discussed in
\cite{EMN}. The reason is that in this case $\mu$ can not be chosen
as a free parameter and for $\beta_m\in\mathbb{Z}_{-}$ the resulting
constraints have no solutions $\alpha_m\in\mathbb{R}$. } In
particular, the parameters $\mu$ and $\nu$ determining the form of
the K\"{a}hler potential are not fixed which considerably extends
the class of allowed K\"{a}hler potentials. They can not be chosen
completely at will however, since one needs to see to that real
solutions for the conditions on $\alpha_m$ and $\beta_m$ in Table
$(1)$ exist. A necessary condition for this is to require
$|\delta|\leq|\gamma|^{3/2}$, assuming $\beta_m$'s to be negative
integers as suggested by the string theory motivated models.
Moreover, the parameters $\mu$ have to be chosen such that
$\gamma<0$.

It is quite interesting to notice that the class of K\"{a}hler
potentials defined by Eq.~(\ref{fullkahler}) and Table $(1)$
includes for example the simple logarithmic form
\beq
\label{kahler}
  K=-{\rm
  ln}\Big(\prod_m(h_m+h_m^*)^{-\beta_m}-\kappa\prod_m(h_m+h_m^*)^{\alpha_m-\beta_m}\p^2\Big)\
  ,
\eeq
where the parameters are now subject to the constraints in Table
$(2)$.
\begin{table}[h!]
\begin{center}
\label{table2}
\caption{The constraints on the parameters of the K\"{a}hler
potential Eq.~$(2.2)$ implied by the flatness of the inflaton
potential.}
\vspace{12pt}
\begin{tabular}{|c||c||c||c|}
\hline\rule{0pt}{3ex}
$\beta=\sum\beta_m$ & $\alpha=\sum\alpha_m$ &
$\gamma=\sum\frac{\alpha_m^2}{\beta_m}$ &
$\delta=\sum\frac{\alpha_m^3}{\beta_m^2}$\\
\hline
$-7$ & $~~0$ & $-\frac{5}{4}$ & $\delta$\\
\hline
$-7$ & $-\frac{25}{9}$ & $-\frac{145}{81}$ &
$-\frac{985}{729}$\\
\hline $-11$ & $-\frac{1}{9}$ & $-\frac{89}{81}$ &
$-\frac{721}{729}$ \\
\hline $-11$ & $-4$ & $-\frac{17}{8}$ &
$-\frac{91}{64}$\\
\hline
\end{tabular}
\end{center}
\end{table}
One can check that solutions for these constraints do exist. As an
example, in the case $\beta=7$, $\alpha=0$ the constraints in Table
$(2)$ are satisfied for a choice
\beq
\beta_m=-1,~\alpha_1=1,~\alpha_2=\alpha_3=\alpha_4=\alpha_5=-\frac{1}{4},
~\alpha_6=\alpha_7=0\ .
\eeq

Both the logarithmic K\"{a}hler potentials Eq.~(\ref{kahler}) and
the more generic forms Eq.~(\ref{fullkahler}) bear some resemblance
to the results appearing in various string theory compactifications.
Up to the quadratic part, the form of Eq.~(\ref{kahler}) is
encountered e.g. in Abelian orbifold compactifications of the
heterotic string theory \cite{abelorbi} and in intersecting D-brane
models \cite{stringmodels}. Logarithmic K\"{a}hler potentials on the
other hand are obtained e.g. in large radius limit of Calabi-Yau
compactifications and also in no-scale supergravity models
\cite{noscale} although the results are not precisely of the form
given in Eq.~(\ref{kahler}). However, it is interesting even in its
own rights that the saddle point condition of the MSSM inflation is
satisfied to the required extraordinary precision with K\"{a}hler
potentials Eq.~(\ref{kahler}) that can be expressed in terms of a
single natural function.


\section{The supergravity corrections to inflation}

The field $\phi$ parameterizing the flat direction has a
non-canonical kinetic term due to the form of the K\"{a}hler
potentials Eq.~(\ref{fullkahler}). Instead of using $\phi$ we
therefore switch to the canonically normalized field
\beq
  \varphi\equiv(\kappa\prod_m(h_m+h_m^{*})^{\alpha_m})^{1/2}\phi~(1+\mc{O}(\p^2))\equiv\hat{Z}_2^{1/2}\phi~(1+\mc{O}(\p^2))\ ,
\eeq
that will be interpreted as the inflaton. Provided the conditions in
Table $(1)$ are satisfied, the inflaton potential in the
supergravity models described above identically becomes \cite{EMN}
\beq
\label{Vlead}
  V(\vp)=\frac{4}{15}m_{\varphi}^2\bvpo^2+\frac{16}{3}\frac{m_{\varphi}^2}{\bvpo}(\vp-\bvpo)^3+
  \xi m_{\varphi}^2\bvpo^7(\vp-\bvpo)+\ldots\
  ,
\eeq
in the vicinity of the point
\beq
\label{cs}
  \bvpo\equiv\hat{Z}_2^{1/2}\bpo\equiv\Big(\frac{m_{\varphi}}{\sqrt{10}\lambda_{\varphi}}\Big)^{1/4}\
  .
\eeq
Here $m_{\varphi}=m\hat{Z}_2^{~-1/2},
\lambda_{\varphi}=\lambda\hat{Z}_2^{~-5/2}$ and the explicit
expressions for them are given in the Appendix. The first two terms
in Eq.~(\ref{Vlead}) arise from the leading order part of the
supergravity scalar potential and, due to the constraints in Table
$(1)$, the lowest order non-vanishing supergravity correction is of
the form $\xi m_{\varphi}^2\bvpo^7(\vp-\bvpo)$ and it is small
enough not to spoil the flatness of the potential \cite{EMN}. The
coefficient $\xi$ is determined by the $o(\p^8)$ part of the
K\"{a}hler potential.

Although the small supergravity correction $\xi
m_{\varphi}^2\bvpo^7(\vp-\bvpo)$ does not invalidate the success of
the MSSM inflation, it may still be significant at the early stages
of the inflationary period where the slope arising from the leading
order potential is very small. Indeed, small linear corrections like
this have been considered without any particular supergravity
motivation in \cite{LythDimopoulos,AM} and it has been shown that
they will affect the resulting spectral index. The difference in our
analysis is that the corrections are not arbitrary but arise from
the supergravity model and are thus completely specified. In the
particular supergravity models considered here, the linear term in
Eq.~(\ref{Vlead}) is also the only relevant correction to the
leading order potential since the higher order supergravity
corrections are too small to leave any observable imprints
\cite{EMN}.

The inflationary properties of the potential Eq.~(\ref{Vlead}) can
be straightforwardly analyzed  \cite{LythDimopoulos,AM} using the
standard slow-roll approximation. If the field starts at rest close
to $\bvpo$, there follows a period of inflation with the amplitude
of curvature perturbation given by\footnote{The expressions for
$\xi<0$ are defined as analytical continuations of the $\xi>0$
results. Here this is trivial and amounts to replacing positive
$\xi$'s with negative ones in the end results.}
\baq
\label{zeta}
  \mc{P}_{\mc R}^{1/2}&\approx&\frac{4}{45\sqrt{5}\pi}\frac{m_{\varphi}}{\xi\bvpo^4}\Big[1+{\rm
  tan}^2\Big({\rm
  arctan}\big(\frac{1}{30\sqrt{\xi}\bvpo}\big)-15\sqrt{\xi}\bvpo
  N_{*}\Big)\Big]^{-1}
  \\\nonumber
  &\approx&\mc{P}_{\mc
  R}^{1/2}(\xi=0)\Big(1-75N^2_{*}\xi\bvpo^2
  +\mc{O}(\xi^2\bvpo^4)\Big)\
  ,
\eaq
and the spectral index by
\baq
\label{n}
  n&\approx& 1-60\sqrt{\xi}\bvpo{\rm
  tan}\Big({\rm
  arctan}\big(\frac{1}{30\sqrt{\xi}\bvpo}\big)-15\sqrt{\xi}\bvpo N_{*}\Big)
  \\\nonumber
  &\approx&n(\xi=0)+300N_{*}\xi\bvpo^2+\mc{O}(\xi^2\bvpo^4)\
  \ .
\eaq
By taking $\xi=0$ one recovers the results of \cite{AEGM} for the
MSSM inflation without any corrections. Here $N_{*}\sim50$ is the
number of e-foldings after the observable scales exit the horizon
and we have assumed that the end of inflation is determined by
\beq
\label{seonloppuny}
  |\eta|=\Big|\frac{V''}{V}\Big|\sim 1 ~~\Longrightarrow~~
  \frac{\bvpo-|\varphi_{\rm end}|}{\bvpo^3}\approx\frac{1}{120}\ .
\eeq

The magnitude of the supergravity corrections in Eqs.~(\ref{zeta}),
(\ref{n}) is determined by the term $\xi\bvpo^2$ and assuming
$N_{*}\sim50$ they become significant for
$\sqrt{\xi}\bvpo\gtrsim8\times10^{-4}$. In the supergravity models
considered here one finds $|\xi|\lesssim\mc{O}(100)$ and for the
field values $\bvpo\sim10^{-4}$ typical in the MSSM inflation
\cite{AEGM, AEGJM}, the corrections can thus become important. On
the other hand, the corrections can always be made negligible by
taking the field values to be small enough $\bvpo\lesssim10^{-5}$,
which corresponds to $|\lambda_{\varphi}|\gg1$ in Eq.~(\ref{cs}). In
this limit the results of \cite{AEGM, AEGJM} are thus recovered.

If the soft mass $m_{\varphi}$ is regarded as an adjustable
parameter, the supergravity corrections can be seen as modifications
of the spectral index alone. This is because the amplitude of
perturbations Eq.~(\ref{zeta}) depends explicitly on $m_{\varphi}$
while the spectral Eq.~(\ref{n}) index does not. By slightly
changing the value of $m_{\varphi}$, the amplitude can thus be kept
fixed while varying the spectral index. However, besides the
spectral index, the supergravity corrections also affect the total
number of e-foldings tending to make the inflationary period shorter
for $\xi>0$ \cite{LythDimopoulos,AM}. For $\xi>0$ the total number
of e-foldings is given by
\beq
\label{Ntot}
  N_{\rm tot}\approx\frac{1}{15\sqrt{\xi}\bvpo}\Big({\rm
  arctan}\Big(\frac{1}{30\sqrt{\xi}\bvpo}\Big)-{\rm
  arctan}\Big(\frac{4}{\sqrt{\xi}}\frac{\bvpo-|\varphi_{\rm
  in}|}{\bvpo^4}\Big)\Big)\ ,
\eeq
which becomes strongly dependent on the initial value of the field
$|\varphi_{\rm
  in}|$ when the supergravity corrections get large and, unlike in
the $\xi=0$ case, the initial conditions\footnote{For a discussion
on initial conditions, see \cite{Allahverdi:2007wh}.} can not be
explained by a period of eternal inflation even in principle since
the classical force always overcomes the quantum effects for
$|\xi|\gtrsim10^{-1}|\lambda_{\varphi}|$. Requiring sufficiently
long period of inflation $N_{\rm tot}\gtrsim50$, Eq.~(\ref{Ntot})
yields an absolute upper bound $\sqrt{\xi}\bvpo\lesssim3\times
10^{-3}$ for the allowed magnitude of supergravity corrections and
using Eq.~(\ref{n}) this implies $n\lesssim1$ \cite{AM}. In the next
Section we show that the parameters in the K\"{a}hler potentials Eq.
(\ref{fullkahler}) can easily be chosen such that this condition is
satisfied.


\section{The spectral index}

As a simple example we first discuss the supergravity corrections
that arise from the logarithmic K\"{a}hler potentials defined by
Eq.~(\ref{kahler}) and Table $(2)$. In this case the supergravity
corrections and in particular the parameter $\xi$ in
Eq.~(\ref{Vlead}) are completely determined since the K\"{a}hler
potential is known to all orders in $\p$. Using standard
supergravity formulae it is then straightforward to work out the
explicit expressions for $\xi$ in each of the cases of Table $(2)$,
see the Appendix. The results are shown in Table $(3)$.
\begin{table}[h!]
\begin{center}
\caption{The coefficient $\xi$ of the lowest order non-vanishing
supergravity corrections arising from the K\"{a}hler potentials
defined by Eq.~(2.4) and Table $(2)$. Here
$\epsilon\equiv\sum\alpha_m^4/\beta_m^3$ and for simplicity we have
given the expressions for $\xi$ to the precision of two digits.}
\vspace{12pt}
\begin{tabular}{|c||c||c||c||c|}
\hline\rule{0pt}{3ex}
$\beta=\sum\beta_m$ & $\alpha=\sum\alpha_m$ &
$\gamma=\sum\frac{\alpha_m^2}{\beta_m}$ &
$\delta=\sum\frac{\alpha_m^3}{\beta_m^2}$ & $\xi$\\
\hline
$-7$ & $~~0$ & $-\frac{5}{4}$ & $\delta$ & $0$\\
\hline
$-7$ & $-\frac{25}{9}$ & $-\frac{145}{81}$ &
$-\frac{985}{729}$ & $-7\times(1.16+\epsilon)$\\
\hline $-11$ & $-\frac{1}{9}$ & $-\frac{89}{81}$ &
$-\frac{721}{729}$ & $-0.25\times(1.00+\epsilon)$\\
\hline $-11$ & $-4$ & $-\frac{17}{8}$ &
$-\frac{91}{64}$ & $-4.8\times(1.16+\epsilon)$\\
\hline
\end{tabular}
\end{center}
\end{table}
Assuming $\beta_m$ to be negative integers, the constraints in Table
$(2)$ imply $\epsilon\equiv\sum\alpha_m^4/\beta_m^3\sim-1$ which
yields an estimate $|\xi|\lesssim1$. As discussed above, the
supergravity corrections to the spectral index Eq.~(\ref{n}) become
significant for $\sqrt{\xi}\bvpo\gtrsim8\times10^{-4}$ and for
$|\xi|\lesssim1$ this requires $\bvpo\gtrsim10^{-3}$. For the
typical field values of the MSSM inflation $\bvpo\sim10^{-4}$ the
corrections are thus negligible and we recover the standard result
$n\sim0.92$ for the spectral index of the MSSM inflation. In the
case $\beta=-7$ and $\alpha=0$ of Table $(2)$ this actually holds
for any field values since the coefficient $\xi$ vanishes
identically.

To discuss the supergravity corrections with the more generic
K\"{a}hler potentials defined by Eq.~(\ref{fullkahler}) and
Table~$(1)$, we first need to determine the potential up to $\p^8$.
The most natural extension of Eq.~(\ref{fullkahler}) is to write
\baq
\label{k8}
K&=&\sum_m\beta_m{\rm
ln}(h_m+h_m^{*})+\kappa\prod_m(h_m+h_m^{*})^{\alpha_m}\p^2+\mu\Big(\kappa\prod_m(h_m+h_m^{*})^{\alpha_m}\Big)^2\p^4
+\nonumber\\&&
\nu\Big(\kappa\prod_m(h_m+h_m^{*})^{\alpha_m}\Big)^3\p^6+\rho\Big(\kappa\prod_m(h_m+h_m^{*})^{\alpha_m}\Big)^4\p^8+\mc{O}(\p^{10})\
,
\eaq
where $\rho$ is a free constant. The coefficient $\xi$ can again be
straightforwardly computed in the different cases of Table $(1)$ and
the result will be of the form $\xi=\xi(\mu,\nu\rho,\epsilon)$,
where $\epsilon\equiv\sum\alpha_m^4/\beta_m^3$. The explicit
expressions are given in the Appendix and by substituting them into
Eq.~(\ref{n}) one readily finds the resulting spectral index.

Just like with the simple logarithmic K\"{a}hler potentials
discussed above, $\xi$ vanishes identically in the case $\beta=-7,
~\alpha=0$ of Table $(1)$ and in this particular case the
supergravity corrections are thus absent. However, the situation is
more complicated in the other cases of Table $(1)$ as can be seen in
Fig. $(1)$ below.
\begin{figure}[!h]
\hspace{-1.8 cm}
\includegraphics*[height=17 cm, width=18 cm]{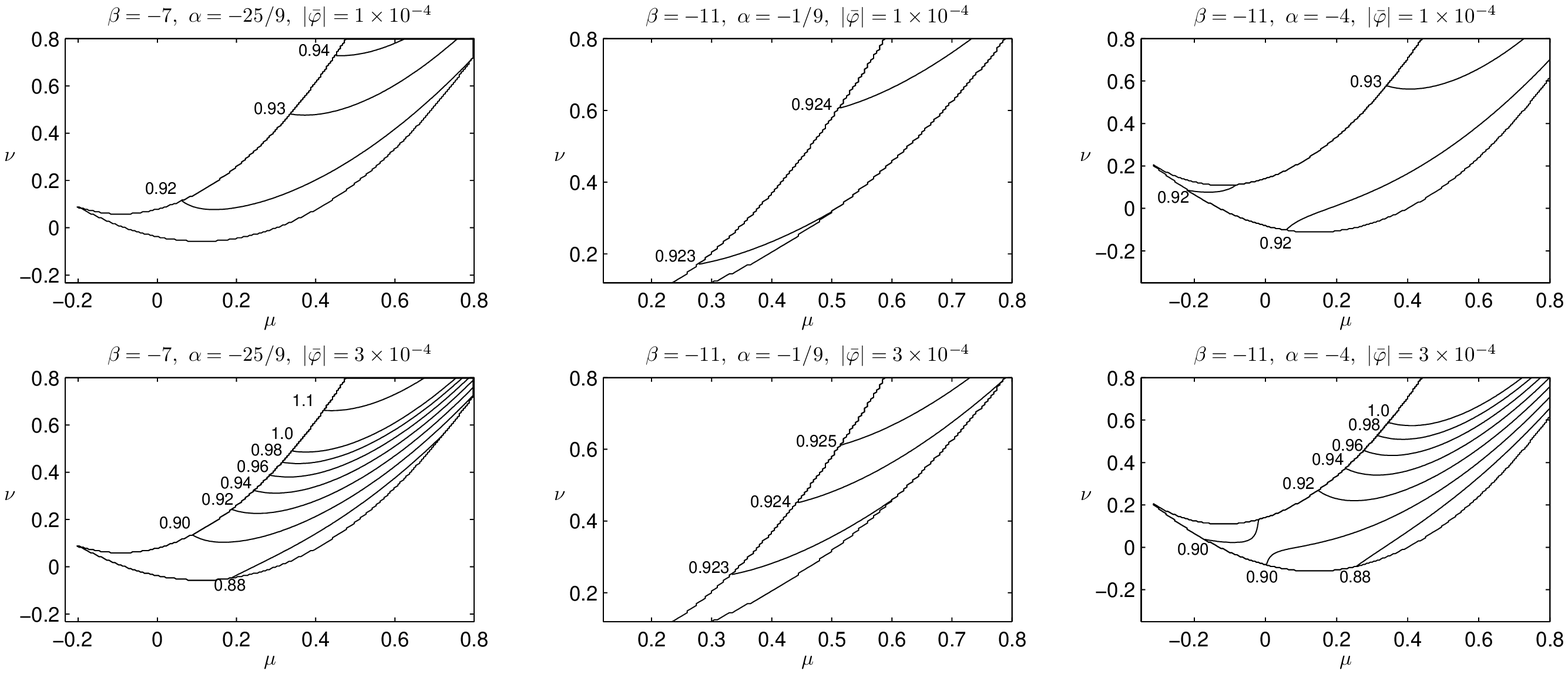}
\vspace{-6 cm}
\caption{The values of the spectral index in the different cases of
Table $(1)$. The case $\beta=-7,\alpha=0$ is trivial with $n=0.923$
and is thus not included. The curves with labels denote countour
lines while the boundary curves describe the region where the
necessary conditions for the existence of solutions for Table $(1)$
are satisfied. Here the parameters $\rho$ and $\epsilon$ in the
K\"{a}hler potential Eq.~(4.1) are chosen as $\rho=1/4$,
$\epsilon=\gamma^2$.
 }
\end{figure}
There the dependence of the spectral index Eq.~(\ref{n}) on the
parameters of the K\"{a}hler potential Eq.~(\ref{k8}) is illustrated
for different field values $\bvpo$. In Fig. $(1)$ we have shown only
the values of $\mu$ and $\nu$ for which the necessary conditions for
the existence of solutions for the constraints in Table $(1)$,
discussed in Section 2, are satisfied. We have also fixed the
parameters $\epsilon$ and $\rho$ appearing in the expression of
$\xi$, but changing their values will not significantly alter the
qualitative behaviour of the spectral index.

Fig. $(1)$ clearly shows that the supergravity corrections typically
tend to bring the spectral index above the value $n\sim0.92$ that
corresponds to the MSSM inflation without any corrections. In the
case $\beta=-11,~\alpha=-1/9$ of Table $(1)$ the corrections are
very small for all reasonable field values and we effectively
recover the value $n\sim0.92$ like in the case $\beta=-7,~\alpha=0$.
In the other cases of Table $(1)$, the corrections are larger and
the range of possible values of the spectral index is highly
dependent on the field value $\bvpo$. For a typical choice
$\bvpo=1\times10^{-4}$ shown in the upper panel of Fig. $(1)$ the
spectral index depends rather weakly on the parameters $\mu$ and
$\nu$ of the K\"{a}hler potential and in the region shown in Fig.
$(1)$ we find $n\sim0.92~...~0.94$. For larger field values the
corrections rapidly become larger and the parameters $\mu$ and $\nu$
need to be chosen more carefully in order to obtain a spectral index
consistent with observations. This is demonstrated in the lower
panel of Fig. $(1)$ for $\bvpo=3\times10^{-4}$. On the other hand,
as discussed in Section 3, for small enough field values
$\bvpo\lesssim10^{-5}$ the supergravity corrections become
negligible and we recover $n\sim0.92$ in all the cases of Table
$(1)$.


\section{Conclusions}

In this work we have discussed the supergravity origin of the MSSM
inflation \cite{AEGM, AEGJM} extending the analysis of \cite{EMN}.
We have shown that the MSSM inflation can be realized in
supergravity models with K\"{a}hler potentials of the simple form
\baq
\label{conc}
K&=&\sum_m\beta_m{\rm
ln}(h_m+h_m^{*})+\kappa\prod_m(h_m+h_m^{*})^{\alpha_m}\p^2+\mu\Big(\kappa\prod_m(h_m+h_m^{*})^{\alpha_m}\Big)^2\p^4
+\nonumber\\&&
\nu\Big(\kappa\prod_m(h_m+h_m^{*})^{\alpha_m}\Big)^3\p^6+\mc{O}(\p^8),\
\eaq
that at least up to the quadratic part closely resemble the results
found in various string theory compactifications \cite{abelorbi,
stringmodels}. The flatness of the inflaton potential is a natural
outcome of such supergravity models provided the parameters
$\beta_m$ and $\alpha_m$ in the K\"{a}hler potential are
appropriately chosen, see Table $(1)$ in the text. However, unlike
in the result found in \cite{EMN}, we have shown that it is not
necessary to completely fix the coefficients $\mu$ and $\nu$. This
considerably extends the class of allowed K\"{a}hler potentials and
thus increases the possibility to find realistic supergravity models
that would yield the MSSM inflation. We wish to emphasize though
that in considering the MSSM inflation \cite{AEGM, AEGJM} driven by
a single degree of freedom, we are implicitly assuming the moduli
fields of the supergravity model to be stabilized by some mechanism
before the beginning of inflation. This represents a non-trivial
assumption and should be discussed separately in the context of any
realistic model to make the analysis complete \cite{Lalak}.

We have also examined the possibility that the underlying
supergravity model would not yield exactly the MSSM inflation
proposed in \cite{AEGM,AEGJM} but a slightly modified model of the
type \cite{LythDimopoulos, AM}. In this case the supergravity
corrections cause small deviations from the saddle point condition
of the MSSM inflation and thus affect the inflationary predictions,
mainly the spectral index. The magnitude of the corrections depends
both on the parameters in the K\"{a}hler potential and on the field
value $\bvpo$ but they typically tend to bring the spectral index
above the value $n\sim0.92$ that corresponds to the MSSM inflation
without any corrections \cite{AEGM,AEGJM}. As an example, for a
natural choice $\bvpo=1\times10^{-4}$ one finds $n\sim0.92~...~0.94$
if the coefficients $\mu$ and $\nu$ in the K\"{a}hler potential are
taken to be less than unity. The range of possible values becomes
larger for larger field values but it is still possible to choose
the K\"{a}hler potential such that the spectral index is consistent
with observations. If the field values are small enough
$\bvpo\lesssim10^{-5}$ the corrections become negligible and we
always recover the result $n\sim0.92$.

The results obtained here and in \cite{EMN} suggest that it might be
possible to realize the MSSM inflation naturally in reasonable
supergravity models. The K\"{a}hler potential certainly needs to be
chosen in specific manner but there is no need for excessive
fine-tuning. The slightly too small spectral index of the original
model of the MSSM inflation \cite{AEGM,AEGJM} may also be easily
cured as discussed above. It would be an interesting subject of
future research to see if these conclusions will change when the
stabilization of the moduli fields and the radiative corrections are
properly taken into account.


\acknowledgments{The author wishes to thank Kari Enqvist, Jaydeep
Majumder and Lotta Mether for useful discussions. The author is
supported by the Graduate School in Particle and Nuclear Physic.
This work was also partially supported by the EU 6th Framework Marie
Curie Research and Training network ``UniverseNet''
(MRTN-CT-2006-035863) .}


\appendix


\section{The supergravity scalar potential}

The supergravity scalar potential is written as
\beq
\label{scalar}
  V=e^K|W|^2\Big(K^{M\N}(K_MK_{\N}+\frac{W_MW^*_{\N}}{|W|^2}+K_M\frac{W^*_{\N}}{W^*}+
  K_{\N}\frac{W_M}{W})-3\Big)\ ,
\eeq
where $K^{M\N}$ is the inverse of the K\"{a}hler metric $K_{M\N}$,
and the lower indices denote derivatives with respect to fields. The
potential for the flat direction $\phi$ is found by substituting the
K\"{a}hler and superpotentials, given by Eqs.~(\ref{k8}) and
(\ref{w}) respectively, into Eq.~(\ref{scalar}).

If the conditions in Table $(1)$ are satisfied and the hidden sector
dependent parts of the superpotential are treated as constants, the
potential can be expanded as in Eq.~(\ref{Vlead}) and the explicit
expressions for $m_{\varphi}$, $\lambda_{\varphi}$ in
Eqs.~(\ref{Vlead}), (\ref{cs}) read
\baq
\label{m}
  m^2_{\varphi}&=&2e^{\hat{K}}|\hat{W}|^2(\alpha - \beta-2)
  \\
  \label{lambda}
  \lambda_{\varphi}&=&e^{\hat{K}/2}\hat{Z_2}^{-3}|\hat{\lambda}_6|\ ,
  \\
\eaq
where we have denoted
\baq
\label{k}
\hat{K}&\equiv&\sum_m\beta_m{\rm
ln}(h_m+h_m^{*})
\\
\label{z}
\hat{Z}_2&\equiv&\kappa\prod_m(h_m+h_m^{*})^{\alpha_m}\ .
\eaq
The coefficients $\xi$ in Eq.~(\ref{Vlead}) in the four different
cases of Table $(1)$ are given by
\baq
\label{xi1}
  \xi_1&=&0
  \\
  \label{xi2}
  \xi_2&=&-\frac{328637}{1509030}-\frac{26674}{3105}\mu^2-\frac{631648}{1035}\mu^3+\frac{376}{69}\nu
  \\
  \nonumber
  &&
  + \mu(-\frac{1154}{1215} + \frac{12864}{23}\nu)
  -96\rho-7\epsilon
  \\
\label{xi3}
  \xi_3&=&\frac{370583}{62329500} - \frac{21457}{64125}\mu^2 - \frac{491524}{21375}\mu^3 +
\frac{484}{1425}\nu
  \\
  \nonumber
  && +
  \mu(-\frac{25921}{577125} + \frac{10434}{475}\nu)
  - \frac{21}{5}\rho-\frac{1}{4}\epsilon
  \\
\label{xi4}
  \xi_4&=&-\frac{8649}{40000} -\frac{15483}{2500}\mu^2 - \frac{263647}{625}\mu^3 +
\frac{516}{125}\nu\nonumber\\&& +
    \mu(-\frac{6909}{10000} + \frac{48528}{125}\nu) - \frac{336}{5}\rho -
    \frac{24}{5}\epsilon
\eaq
where the subindices refer to the rows of Table $(1)$.

\end{document}